\documentclass[revsymb, preprint]{revtex4}
\usepackage{amsmath,amsfonts,amssymb,graphicx}
\usepackage{float}
\floatstyle{plain}
\restylefloat{figure}
\restylefloat{table}
\begin{document}
\title{Chiral Dirac Fermions on the Lattice using Geometric Discretisation}
\author{
Vivien de Beauc\'e and Samik Sen,}
\email[e-mail:]{debeauce@maths.tcd.ie, samik@maths.tcd.ie}
\affiliation{School of Mathematics, University of Dublin, Dublin 2, Ireland.} 
\begin{abstract} 
We propose a discretisation scheme based on the Dirac-K\"ahler formalism (DK)
in which the algebraic relations between continuum operators 
$\{\wedge, \; d, \; \star\}$ are captured by their discrete analogies, 
allowing the construction 
of the relevant projection operators necessary to prevent species doubling. 
We thus avoid the traditional form of species doubling as well
as spectral doubling, which does not occur in the DK setting.
Chirality is also captured, since we have $\star$ from geometric discretisation. Some 
remarks regarding the gauging of the theory are made.
\end{abstract}
\maketitle
%11.15.Ha, 11.30.Rd, 2.40.Sf

\section{Introduction}
The lattice provides a regularisation of continuum quantum chromodynamics 
\cite{Wilson} which has played a central role in the study of non-perturbative 
effects such as quark confinement. However, problems arose which seem, at 
least at first sight, to be inherent to the lattice.

Putting fermions on the lattice naively leads to degeneracy, as can be seen
from the dispersion relations obtained. When Wilson removed this
degeneracy, it may have seemed that the loss of chirality was 
perhaps coincidental but subsequent work \cite{KS, NN}, based on 
topological arguments, found that this was not the case.

The argument, proposed by Nielsen and Ninomiya (NN), shows that there 
are, under reasonable assumptions such as exact conservation of 
discrete valued quantum numbers 
(lepton number $Q$) and locality, an equal number of species of left and
 right-handed neutrinos which means that you cannot have single fermions since
 they occur in pairs. Handedness is then taken into account by 
considering the homotopy class of maps from closed surfaces, encircling the 
degeneracy points $ \omega_{deg}$ embedded in the Brillouin zone into the 
space of rays formed by the N fundamental fermion components 
$\mathbb{CP}^{N-1}$.
The degenerate fermionic states, $\mid \omega >$, live in this space
and satisfy 
\[ H \mid \omega > = \omega_{deg} \mid \omega >.\]  

The Dirac-K\"ahler (DK) formulation \cite{kahler}, which is differential 
geometric \cite{NS, Lang, Flanders} in nature, 
is advantageous since it gives rise to a non-degenerate energy spectrum in the
discrete formulation, which means that the NN theorem is not applicable \cite{becher}.
With this in mind a discrete DK theory \cite{KSuss,BJ} can be expressed,
using ideas from algebraic topology to provide a natural discrete analogue to 
differential geometry. Becher and Joos \cite{BJ}
showed how doubling still arises in this theory after reduction, even though topology is 
captured and NN is not applicable \footnote{The no-go theorem of Nielsen-Ninomiya can also 
be avoided using Ginsparg-Wilson \cite{GW} methods; namely domain wall and 
overlap fermions \cite{Creutz}.}, as the discrete operations used do not 
satisfy the desired properties.

The algebraic properties of the discrete analogies of the operators 
$\{\wedge,\; d,\; \star \}$
are used to describe spinors and the action of the Clifford algebra on them.
The latter has a differential geometric analogue through the introduction of
the Clifford product (CP), both in the continuum and on the lattice.
Its action, denoted by $\vee$, is a combination of the operations in the triple 
$\{\wedge,\; d,\; \star \}$ and plays an important role in the construction.

We use the DK approach with geometric discretisation
(GD) \cite{Us} as our discrete differential geometry.
GD uses discrete analogies to continuum objects and operators, capturing 
Stokes' theorem and the Hodge decomposition theorem, 
in a similar fashion to Dodziuk \cite{Dodziuk}. It also possesses a discrete
Hodge star operator by using a subdivided space in which both the original and dual lattice are
contained, which was previously not available. The discrete 
wedge enables us to impose the conditions needed to
relate the Dirac-K\"ahler equations to the Dirac equation (DE)
\cite{GSW}, which was the problem faced by Becher and Joos \cite{BJ}, and 
the discrete Hodge star allows us to deal with chirality as was 
discussed by Rabin \cite{rabin}.

To summarise:
\begin{itemize}
\item We use the DK approach and so do not suffer from spectral degeneracy.
\item In our framework we use:
\begin{itemize}
\item The GD wedge, which means that we can impose subsidiary conditions 
to avoid species doubling.
\item The GD Hodge star, which means that we have chirality. 
\end{itemize}
\end{itemize}

In this work, we strengthen the relation between the lattice and 
the continuum in order to propose a more satisfactory lattice regularisation 
of the Chiral theory.

We begin with a brief review of the DK equation in section 2. We then look at how this
is dealt with discretely in section 3; first, using standard methods
where the wedge fails, and then using GD where it does not. We address some issues regarding the gauging of the free theory in section 4.

\section{The Dirac-K\"ahler Equation}
In the language of differential forms the Laplacian is given by
\[ \triangle = d\delta+\delta d.\]
Thus we note that
\[ -(d-\delta)^2= \triangle, \]
from which we obtain the Dirac-K\"ahler equation (DKE) \cite{kahler, kruglov}
\[ i(d-\delta +m)\psi = 0, \]
 the solution of which has the following the functional form
\[ \psi=1+f_\mu dx^\mu + \frac{1}{2!}f_{\mu \nu }dx^\mu \wedge dx^\nu+
\frac{1}{3!}f_{\mu \nu \lambda}dx ^\mu \wedge dx^\nu \wedge dx^\lambda+
f_{0123}dx^0 \wedge dx^1\wedge dx^2 \wedge dx^3.\]

In order to represent spinors in this language, we introduce the Clifford 
product, which acts on the space of 
inhomogeneous differential forms and satisfies the following
relations:
\begin{eqnarray}
\label{Eqn:CPrules}
1 \vee 1 &=& 1, \\
1 \vee dx^\mu &=& dx^\mu \vee 1 = dx^\mu, \nonumber \\ 
dx^\mu \vee dx^\nu &=& g^{\mu\nu}\cdot 1 + dx^\mu \wedge dx^\nu, \nonumber 
\end{eqnarray}
where $g^{\mu\nu}$ is the Euclidean metric.
Through the identification 

\begin{equation}
\label{id}
dx^\mu \vee \mapsto \gamma^\mu, 
\end{equation}
which arises from representation theory, we relate the differential forms 
under the CP to the algebra of gamma matrices. It is now immediate that
\[ \{ \gamma^\mu,\gamma^\nu\}=2g^{\mu\nu}.\]

The Dirac field then belongs to a 16 dimensional representation 
of the algebra of gamma matrices. Furthermore all the representations
of the complex Clifford algebra of the 4-dimensional Euclidean space
can be decomposed into 4-dimensional irreducible representations \cite{reptheory};
these being equivalent to those generated by the standard gamma matrices.

As we have introduced the Clifford product, an important observation due to
Susskind is now in order, namely, that the DKE is invariant under the
group $SU(4)$, referred to as a global flavour symmetry, which acts by 
right action of the Clifford product with a constant differential $U$.
That is,

\begin{equation}
\label{associativity}
\left( d - \delta +m \right) \left( \Phi \vee U \right) = \{ \left(d - \delta +
m \right) \Phi\} \vee U = 0, 
\end{equation}
this being established through the identification $d - \delta \mapsto dx^{\mu} 
\vee \partial_{\mu}$ and use of the associativity of $\vee$ which is known to 
hold in the continuum.

We can thus decompose the 16D space of differential forms $\mathcal{D}=\{\Phi \}$ into
4-dimensional invariant subspaces
\[ \mathcal{D}=\bigoplus_{b=1}^4 \mathcal{D}^{(b)}, \]
on which the DKE implies the DE.

One can construct a collection of projection operators $P^{(b)}$
mapping forms onto the irreducible subspaces $\mathcal{D}^{(b)}$, where
\begin{equation}
 \Phi \vee P^{(b)} = \Phi,
\label{Eq:subsidiary}
\end{equation}
if $\Phi \in \mathcal{D}^{(b)}$. We also know that $\Phi \vee P^{(b)}$ is  
 a solution of the DKE from the argument given above regarding the global $SU(4)$ symmetry.
With this subsidiary condition \eqref{Eq:subsidiary}, the DKE for fixed
b is equivalent to the DE.

From \eqref{Eqn:CPrules} we can see that 
$d - \delta = dx^{\mu} \vee \partial_{\mu}$.  Considering 
\[
\label{g}
\Phi^{(b)} = \Phi \vee P^{(b)},
\]
where $\Phi$ is a solution of the DKE,  we get
\[
\label{h}
\left( d - \delta +m \right) \Phi^{(b)} \mapsto \left(\gamma^{\mu}
\partial_{\mu}
 +m \right) \Phi^{(b)},  
\]
leading to the Dirac equation.

We note that the DKE can also be expressed using the set of equations
\[ i(d\omega^{(p-1)}-\delta \omega^{(p+1)})=-m\omega^{(p)}, 
\,\,\,\, p=\{0,1,\dots,4\}. \]
These are related. For instance, the
$p=0$ and $p=4$ equations are equivalent; as can be shown using 
$\star\star =I$ and $\delta=\star d\star $. A solution to the $p=0$ equation living in one of the $\mathcal{D}^{(b)}$ is then a solution of the $p=4$ equation (in the same $\mathcal{D}^{b}$). If we did not have 
$\star\star=I$ and $\delta=\star d\star$ then the
relations between the equations would not be captured and degeneracy would arise.

In this language, taking $\gamma^5=\gamma^0\gamma^1\gamma^2\gamma^3$ 
and with the identification $dx^u \vee \mapsto \gamma^\mu$ it is found that
$\gamma^5$ corresponds, up to sign, to $\star$ ; $(\gamma^5)^2=1$ 
while $\star \star=(-1)^p$.
Thus the Hodge star operator plays a central role in formulating
chiral fermions \cite{rabin}.

Before we address the discretisation of this theory, it is worth looking at 
the definition of the Clifford product in the continuum, which we express in 
the form
\begin{equation}
\label{po}
\Phi \vee \Psi = \sum_{p \geq 0} \frac{{\rm sign} \left(p\right)}{p !} \left(\eta^p ( e
_{\mu_{1}} \rfloor \ldots e_{\mu_{p}} \rfloor \Phi) \right) \wedge \left( e^{\mu
_1} \rfloor \ldots e^{\mu_{p}} \rfloor \Psi \right), 
\end{equation}
where { \it ${\rm sign} \left(p\right) = (-1)^{\frac{p(p-1)}{2}}$ \it} and $\eta^p \omega^{r} = (-1)^{r} \omega^{r}$. We note that the operations 
of contraction $\rfloor$ and that of wedging are both used. 

\section{The Dirac-K\"ahler Equation on the Lattice}

After having stressed the importance of the subsidiary conditions above, we now
 investigate 
their discrete counterpart \footnote{Other approaches of interest are given in \cite{vaz, issaku}.}. We will first explain the nature of the problem by 
describing the usual method and then propose our alternative. Again, this will 
turn out to be linked to the algebraic relations of the discrete operators; 
particularly of the wedge $\wedge$.

At this point it is useful to note that the (continuum) subsidiary conditions can
be given the form of a group property
known as the reduction group denoted ${\cal R}$ generated by
$\tau=idx^1\wedge dx^2$ and $\epsilon=dx^1\vee dx^2\vee dx^3 \vee dx^4$
under $\vee$-multiplication.

On the lattice, we represent $p$-forms as $p$-cochains. Thus the Dirac
fields are represented by inhomogeneous cochains. We now review the 
construction of the discrete analogue of the reduction group.

\begin{figure}[H]
\label{fig:T}
{\par\centering \includegraphics{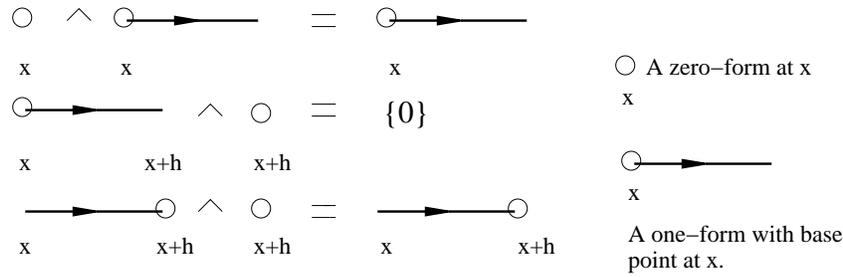} \par}
\caption{Role of the base point to the discrete wedge}
\end{figure} 

In that approach, where the cup product is taken as the discrete wedge,
the operation is performed at a base point $x$.
The left and right cochain arguments of the wedge must have the
same base point, as illustrated by the above FIG. 1.
%[figure shows that if right then works, otherwise doesn't]

In the continuum $x \wedge y$ is equal to $y \wedge x$ up 
to sign \footnote{The discrete wedges of Birmingham-Rakowski \cite{starproduct} 
and
Albeverio-Sch\"{a}fer \cite{albeverio} satisfy this property also.} 
which is no longer the case on the lattice; unless we translate the right 
cochain in such a way that its base point coincides with the other, enabling us to wedge cochains which do not have the same base point. 

In the continuum we have
\[ dx^1\vee (dx^1\wedge dx^2) = 1 \wedge dx^2  = dx^2 \]
because we remove any $dx^i$'s which two arguments have in common 
before wedging them, when taking their Clifford product.

So,
\[ (dx \wedge dy) \vee (dx \wedge dy) = 1 \wedge 1 =I\]
whose discrete analogue is
\[ (dx\wedge dy) \vee T (dx\wedge dy)= T_{e}I,\]
where translation operation $T_e$ shifts simplices $\sigma$ and $\eta$ so that
$\sigma \wedge \eta$ and $\eta \wedge \sigma$ have common support.

It is then found that \cite{BJ}:
\begin{align*}
&\tau^{2}= T_{- \left(e_{1} + e_{2} \right)}, \; \; \epsilon^{2} = T_{-
\left(e_{1} + e_{2} + e_{3} + e_{4} \right)}; \\
& \tau \vee \epsilon = \epsilon \vee \tau = T_{e_{1} + e_{2}} \tau \epsilon,
\end{align*}
thus supplementing the group $ \mathcal{R}$ with translation elements.
This means that the discrete reduction group does not close and 
it is not possible to impose the subsidiary conditions.

Note that the presence of translation operators in the discrete analogue of the
 reduction 
group is unavoidable in this formalism. One could consider the space resulting 
from quotienting translation elements, but this would lead to the lattice 
being identified as one unique point.

\begin{figure}[H]
\label{fig:S}
{\par\centering \includegraphics{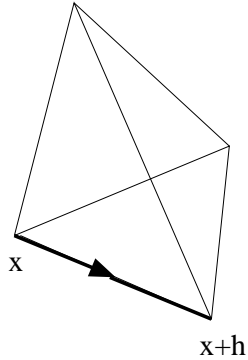} \par}
\caption{Discrete wedge in GD: the higher dimensional region, here a 
tetrahedron gives common support to the vertex at $x$, the vertex at $x+h$ and 
to the edge $[x, \; x+h]$.}
\end{figure} 

In geometric discretisation (GD) \cite{Us} we map any $p$-cochains, $\sigma^p$, to $p$-forms,
$\omega^p$, using the Whitney map $W$ and wedge these, before mapping the result 
back to the lattice, using the de Rham map $A$ given by 
\begin{equation}
A(\omega^p) = \sum_i \left(\int_{\mid \sigma^p_i \mid} \omega^p \right)
[ \sigma^p_i],
\label{equation:derham}
\end{equation}
where $AW=I$ which makes the Whitney map the right-inverse of $A$ and in the standard simplex coordinate system is given by 

\begin{equation}
\label{whitneymap}
W ([v_{0} \ldots v_{r}]) = r ! \sum_{i=0}^{r}  (-1)^{i} \mu_{i} d\mu_{0} \wedge \ldots \wedge \hat{d\mu_{i}} \wedge \ldots \wedge d\mu_{r}, 
\end{equation}

and the functions $\{\mu_{i}, i \in [0, \, r] \}$ are the so called barycentric coordinate functions.

\begin{figure}[H]
\label{fig:U}
{\par\centering \includegraphics{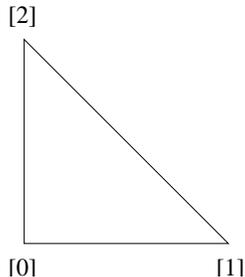} \par}
\caption{The standard simplex}
\end{figure} 

In the case of the triangle of FIG. 3, the barycenter coordinates are given by

\begin{equation}
\label{themus}
\mu_i = 
\begin{cases}
 1 - x - y, \; \text{for} \; i=0, \\
x,  \; \text{for}\; i=1, \\
y,  \; \text{for} \; i=2.
\end{cases}
\end{equation}

It is then immediate to check that 

\begin{equation}
\label{stokes}
d W^{K} ( [v_{0} \ldots v_{r}]) = W^{K} d^K [v_{0} \ldots v_{r}], 
\end{equation}
displaying how the simplicial map $d^K$ is tied to its continuum analogue $d$.
Starting with \eqref{themus}, and applying \eqref{whitneymap}, one obtains the so called Whitney elements which form the basis list of differential forms captured by the cochain space under $W$\footnote{The complex dual to simplices is not simplicial, but hexagonal, for example, for which we do not have a Whitney map. We thus favour the hypercubic lattice whose dual is also hypercubic, and for which a Whitney map has been proposed \cite{cubic}.}

We then define  the discrete wedge of any two given simplices $\sigma$ and $\eta$ living in the complex $K$ as 

\begin{equation}
\label{discretewedge}
\sigma \wedge^K \eta = A^{K}(W^{K}(\sigma) \wedge W^{K}(\eta)),
\end{equation}
which satisfies $x \wedge^K y = \pm y \wedge^K x$ 
with no translation element required. Thus $\tau^2=I$ and the reduction group 
closes. 
We note that the wedge product in our method 
is only associative up to a multiplicative factor which is a function of the degree 
of the cochains being wedged. Given three cochains $\sigma, \; \eta, \; \zeta$ in K, respectively of degrees $p$, $q$ and $r$, the rule is

\begin{equation}
\label{non-assoc}
\sigma \wedge^K (\eta \wedge^K \zeta)= \left(\frac{p+q+1}{r+q+1}  \right)(\sigma \wedge^K \eta) \wedge^K \zeta.
\end{equation}

Before we proceed to discuss the Dirac-K\"ahler equation in our framework, we would like to comment further on the simplicial operations defining GD by giving our discrete exterior derivative $d^K$ and its adjoint $\delta^K$. They are as follows:

\begin{equation}
d^K [v_{0} \ldots v_{r} ] = \sum_{k=0}^{N} [v_{k}v_{0} \ldots v_{r}];
\end{equation}
while,

\begin{align*}
\delta^{K}[v_{0} \ldots v_{r}] &= \star^{L} d^L \star^{K} [v_{0} \ldots v_{r}]\\
           &=\pm \partial_{K} [v_{0} \ldots v_{r}] \\
&= \sum_{k=0}^{N} \left(-1\right)^{k} [v_{0} \ldots \hat{v}_{k} \ldots v_{r}] 
\end{align*}
where $N$ is the number of vertices. Note the role played by the dual complex $L$ in defining $\delta_{K}$.

Example: Consider FIG. 3 again and the zero-cochain 
\begin{align*}
\sigma_{K}^{0}&= A^K ( f)= f([0]) [0] + f([1]) [1] + f([2]) [2] \\
\intertext{Applying $d^K$ we obtain} 
d^K \sigma^0 &= (f([1]) - f([0])) [01] + ( f([2]) - f([1]))[12] +
( f([2]) - f([0]))[02].
\end{align*}

Noting that for $[0]=(0,0)$ and $[1]=(h,0)$ we see that $f([1])-f([0])=f(x+h)
-f(x)$. Also, since we started with a zero-cochain, $\delta \sigma^{0}_{K} = 0$.
The emerging picture for the DK equation in the present framework is that one has a hyper-cubic complex with fields represented by inhomogeneous cochains with constant coefficients in front of simplices such as $[i]$, $[ij]$, $[ijkl]$ and so on. The fact that one has Whitney elements is of importance here in two related ways. First of all, they provide support to the associated form and also are used in the wedge formulae used below.

%% To obtain the general term of degree one from the DK equation we would need to consider the one-form field 

%%\begin{equation}
%%\sigma^{1}_{K} = h ([01]) [01] + h ([12]) [12] + h([02]) [02] 
%%\end{equation}

%%where $h([ij]) = \int_{[ij]} h_x (x, y )dx + h_y (x, y) dy$. Similarly for the two-form

%%\begin{equation}
%%\sigma^{2}_{K} = g ( [012]) [012]
%%\end{equation}

%%which leads to
%%\begin{equation}
%%\delta_K ( g ( [012]) [012]) = g([012])([12] - [02] + [01]).
%%\end{equation} 

%%The DK equation then will read, keeping only the one-cochain term, and acting on the inhomogeneous cochain field $\sigma_K=  \sigma^{0}_{K} + \sigma^{1}_{K} + \sigma^{2}_{K}$ gives

%%\begin{align*}
%%(d_K - \delta_K + m)\; \sigma_K &= (f([0]) - f([1]) - g([012])+ m h([01])[01] \\
%%&+ (f([1]) - f([2]) - g([012]) + m h([12]) )  [12] \\
%%&+ (f([0]) - f([2]) + g([012]) + m h ([02])) [02] 
%%\end{align*}

To compare this with the method described previously, we point out that  
the vertices at $x$ and at $x+h$ have overlapping support on 
the edge $[x, \; x+h]$ (FIG. 2). If one was to specify a base point to be
$x$ or $x+h$, as is done in turn in FIG 1, one would still get the 
correct non-zero answer. It follows that no translation is necessary.

In the argument given above, we assumed that the inhomogeneous differential forms projected 
under the right action of $P^{(b)}$ remain solutions of the DKE. In the 
continuum, this step involved the associativity of the continuum Clifford product. 
As is apparent from \eqref{associativity} the associativity of the discrete Clifford is 
conditional on the discrete wedge being associative. In other methods such 
as \cite{BJ} the discrete Clifford product is non-associative in general 
but does not affect the derivation of \eqref{associativity}. In the 
present framework we point out that we are given the exact expression for the 
$P^{(b)}$ differentials in the discrete theory. Therefore, owing to the fact 
that the Clifford product gives rise to the algebra displayed in \eqref{Eqn:CPrules}, we can assert that
\begin{equation}
\label{er}
\Phi^{'} = \Phi \vee P^{(b)}
\end{equation}
is also a solution of the DKE.

More explicitly, the $P^{(b)}$ are constant differentials, and our discrete wedge \eqref{discretewedge} is exact for constant differentials. Hence provided the lattice field $\Phi^K$ satisfies the discrete DK equation

\begin{equation}
\label{dkek}
(d^K - \delta^K + m ) \Phi^K = 0, 
\end{equation}
the equations

\begin{equation}
(d^K - \delta^K + m ) (\Phi^K \vee^K P^{(b)}) = 0
\end{equation}
follow without requiring associativity of the discrete wedge. To make sense of this last equation in the lattice setting, we recall the formulae for the Clifford product does include the wedge but also contraction \footnote{An approach for contraction and symmetries by one of the authors V de B using GD will appear in a later paper.}. Given a two form for example which has support on a square $[0123]$ say with $[01]$ and $[32]$ opposite edges along the $x$-axis and $[03]$, $[12]$ parallel edges along the $y$-axis. Then contracting the two form represented by $[0123]$ with a $dx$ and using (see Becher and Joos \cite{BJ})

\begin{equation}
P^{(b)} = \frac{1}{4} ( 1 + i \text{sign}(12) dx^1 \wedge dx^2) \vee ( 1 + \text{sign} (1234) . \epsilon)
\end{equation}
in \eqref{er}, we have an analogous lattice operation 

\begin{equation}
e_{x} \rfloor^K [0123] = [03] + [12],
\end{equation}
while $W^K([03])$ and $W^K ([12])$ have respectively images given by $(1-x)dy$ and $x dy$ so the contraction operation at the simplicial level gives back the correct Whitney element of form degree one less. This is what is meant by capturing exactly the right multiplication with the constant differential $P^{(b)}$. It is of interest to compare this approach with the discussion of Rabin \cite{rabin} in which the importance of the discrete operators in the reduction is identified.

On the issue of chirality, our discrete Hodge star maps one space into 
its dual, that is there  
is a doubling of the space of cochains. As is customary in lattice theories we 
then deduce the relation:
\begin{equation}
\label{qwe}
\{ \gamma_{5}, \; D \} = 0.
\end{equation}

In the present DK formalism this is expressed (acting on fields in K, but one can equally swap K and L superscripts) as 

\begin{equation}
\{\star,d-\delta\}^K \equiv \star^{K} (d^K - \delta^K ) + (d^L - \delta^L) \star^K,
\end{equation}
since $\star$ plays the 
role of $\gamma_5$ and
$D=d-\delta$. Using $\star^L\star^K=I$ and $\delta^K=\star^L d^L \star^K$, which are both satisfied in GD, we 
find
\begin{align*}
\star^K (d^K-\star^L d^L\star^K )+(d^L- \star^K d^K\star^L)\star^K&=(\star^K d^K-\star^K\star^L d^L \star^K)+
(d^L\star^K -\star^K d^K\star^L \star^K)\\
&=(\star^K d^K -d^L\star^K)+(d^L\star^K -\star^K d^K)=0.
\end{align*}
Note that $\gamma_5$ is identified with $\star^K$ and $\star^L$, up to some sign factor, since 
$(\gamma_5)^2=1$ where as $\star^L \star^K=\pm 1$. So we always have 
$\star \star =1$ provided we insert the correct sign factor,
leading to the desired result. Again, there are two spaces involved here,  
since $\star^K D^K$ and $D^L\star^K$ both map objects to their dual space,  which is 
also the case in Ginsparg-Wilson \cite{GW} for example.

The presence of the two spaces, $K$ and $L$, is, we claim, not a problem in the 
discretised chiral fermion theory. In order to check this, we note that the  
subsidiary conditions carry through 
under application of $\star$. Namely, if $\omega \in P^{(b)}_{K}$ then $ \star 
\; \omega \in P^{(b)}_{L}$.
Since the theory is well defined as we interchange $K$ and $L$, a space such as
 $P^{(b)}_{L}$ is well-defined. That is, in a self-explanatory notation we can write an $L$-DKE as we were able to write a $K$-DKE in \eqref{dkek}.

Given the inhomogeneous fields $\Phi_{K}$, represented by inhomogeneous cochains 
in the complex $K$, we can now consider the associated action functional $S_K$:
\[ S_K = < \bar{\Phi}_K,(d-\delta+m)\Phi_K>, \]
where (denoting complex conjugation by c) 
\[ \bar{\Phi}_K=\gamma_5 \Phi^c_K=\star^K \Phi^c_K \]
and so belongs to the dual space $L$. It is worth pausing for a moment to appreciate the role played by $\star$ both as $\gamma^5$ and as providing the volume form for integration. 

The inner product $<\cdot, \cdot>:C^p(L) \times C^{n-p}(K) 
\longrightarrow \mathbb{C}$ \cite{Us} is given by 
\[ <\sigma^L, \eta^K> = (\sigma^L, \star^K \eta^K) = 
\frac{(n+1)!}{p!(n-p)!}\int_M W^B(B\sigma^L)\wedge W^B(B\eta^K), \]
where $B$ is the subdivided space containing both $K$ and $L$ and
$B\sigma^L$ is the projection of $\sigma^L$ onto $B$.

The action is of the form 
\begin{equation}
\label{action}
S_{K} \equiv S \left( \Phi_{K} , \; \star^K \Phi_{K}' \right),
\end{equation}
where there is a subtlety since the field $\star^K \Phi_{K}$ lives in $L$. 
Although there is a field in the original and dual complex, one is the
image of the other.
We can repeat the argument starting with the $L$ complex, and define
$S_{L}\left(\Phi_{L} \right)$. This in contrast with methods such as \cite{BJ} in which the volume element is induced by right Clifford multiplication with $\epsilon$. Again, this is due to the close link between the lattice and differential geometry in our method.

Finally we look at the partition function for which  
the action is a direct sum of its  components in the 
invariant spaces $\mathcal{D}^{(b)}$.
\begin{equation}
\label{wer}
S = \sum_{b=1}^{4} S^{(b)},
\end{equation}
which according to the argument above can be readily rewritten as
\begin{equation}
\label{tyu}
S = \sum_{b=1}^{4} S^{(b)}_{K} \oplus S^{(b)}_{L}.  
\end{equation}
Furthermore, $S^{(b)}_{K}$ and $S^{(b)}_{L}$ are functionals respectively of 
the fields $\Phi_{K}$ and $\Phi_{L}$ and their associated representative in the 
dual spaces $L$ and $K$. It follows then that the counting of the fields is correct,
while we obtain the partition function squared given below. 
To see this, we introduce a matrix $M$ determined by taking
\begin{equation}
\label{pq}
\Phi^{T}_{K} M^K \Phi'_{K} = S_K (\Phi_{K}, \; \Phi'_{K} ).
\end{equation}
The entries of $M^K$ are fixed, up to normalisation (volume factors),  by 
\[ M^{K}_{ij}=S_{K}(\sigma_i,\sigma_j). \]
We can fix this by normalising the parameters in the general formula
for the discretised fields,
\[\Phi_K=\sum_i \lambda_i [\sigma_i],\]
from Eq. (\ref{equation:derham}). 
The space of parameters $\lambda_i$ determines the partition function
measure:
\begin{equation}
\label{poi}
Z = \int \; [d \Phi_{K}][d \Phi_K^T] [d \Phi_{L}][d \Phi_L^T]
e^{\; \Phi_{K}^{T} M \Phi_{K} + \Phi_{L}^{T} M \Phi_{L} } = \int\; [d \Phi_{K}][d \Phi_K^T]e^{\; \Phi_{K}^{T} M \Phi_{K}}\int\; [d \Phi_{L}][d \Phi_L^T]e^{\; \Phi_{L}^{T} M \Phi_{L}}
\end{equation}
which splits into two independent parts, one for fields defined on $K$ and the other for 
fields defined on $L$. So, there is no mixing between the two lattices $K$ and $L$.

\section{Remarks on gauging the free fermions}

In the DK picture the global $SU(4)$ symmetry is linked to the specific representation of fermions using the CP, and it has been argued that one could attempt to make it into a local ``flavour'' gauge symmetry.
Here however, one take the inhomogeneous differential forms, and add a gauge 
label to their coefficient turning a given form into a Lie algebra valued form. The central observation is that the group action on the gauge labels should leave the reduced spaces 
$P^{(b)} \Omega$ invariant. In the continuum language one would say that reduction (which is an action of $SU(4)$ on $\Omega$) commutes with gauging the free fermions. 

%% \begin{equation}
%% \label{qoi}
%% \omega = \lambda_{(a), \, \mu} T^{(a)} \; \left(\bigwedge dx^{\mu}\right)
%% \end{equation}

In GD, some freedom is given to us in how one might gauge the fermions, let us choose minimal coupling. As is customary, introduce gauge fields $A_{\mu} \left(x\right)$ and consider the parallel transport operator from the vertex at $x$ to the vertex at $x + \hat{\mu}$:

\begin{equation}
\label{gre}
U_{\mu} \left(x \right) = \exp  [i e a A_{\mu} \left(x\right)].
\end{equation}

The operator $U$ is then inserted in the fermion bilinear term of the free Lagrangian. The invariance of the Lagrangian under gauge transformation follows by the usual manipulation of matrices and traces. In effect, we have coupled a zero-form (in the spinor representation of DK only) gauge field to an inhomogeneous form field. Furthermore, this guarantees that the $\mathcal{D}^{\left(b\right)}$ are left invariant.

%% How the gauging is implemented in practice has to do with the procedures of blocking and thinning \cite{Kovacs} briefly described below.

In the usual Dirac-K\"ahler methods, as described above in geometric language, one diagonalises the free Lagrangian leading to

\begin{equation}
\label{gauge}
L = a^{d} \sum_{r, \, \mu} \bar{\phi} \left(r\right) \frac{1}{2a} \frac{1}{2^{\frac{d}{2}}} [ Tr \, [ \left( \Gamma^{\frac{r}{a}}\right)^{\dagger} \gamma_{\mu} \Gamma^{\frac{r}{a} + \hat{\mu}}] \phi \left( r + a \hat{\mu} \right) - Tr \, [ \left( \Gamma^{\frac{r}{a}} \right)^{\dagger} \gamma_{\mu} \Gamma^{ \frac{r}{a} - \hat{ \mu}} \phi \left( r - a \hat{\mu} \right)]]
\end{equation} 
by means of the transformation of Kawamoto and Smit \cite{kawamoto} (introducing the $\Gamma$ matrices) followed by thinning which amounts to taking the trace leaving one component of $\phi$. At this point, one has the option of blocking the fields  \cite{napoly, Kovacs}; this is done by introducing the so-called block coordinates and by means of a unitary transformation one obtains a Dirac $\otimes$ flavour representation. The gauging is then done by inserting link variables in the Lagrangian \eqref{gauge}.

Of particular interest to us, is the framework of staggered fermions in which the gauging is done after the thinning. One obtains directly:

\begin{equation}
\label{g2}
L^{S} = a^{d} \sum_{r, \, \mu} \bar{ \phi} \left(r\right) \frac{1}{2^{\frac{d}{2}}} Tr \, [ \left( \Gamma^{ \frac{r}{a}}\right)^{\dagger} \gamma_{\mu} \Gamma^{\frac{r}{a}+ \hat{\mu}}] \frac{1}{2a}[ U_{\mu} \left(r\right) \phi \left( r + a \hat{\mu}\right) - U^{\dagger}_{\mu} \left(r- a \hat{\mu} \right) \phi \left( r - a \hat{ \mu} \right)].
\end{equation} 

 The reduction described in this paper which leads to \eqref{action} can be interpreted as the analogue of diagonalising and thinning, we reduce the degrees of freedom to obtain one fermion with $2^{\frac{d}{2}}$ spinorial components described in any given invariant subspace $\mathcal{D}^{(b)}$ by an action of the form \eqref{g2}. However we can also be more general and consider

\begin{equation}
\label{g3}
S_{K} = \langle \bar{ \Phi}_{K}^{\left(b\right)}, \mathbf{U} \,  [ \left( d^{K} - \delta^{K} +m \right) \Phi^{\left(b\right)}_{K}] \rangle ,
\end{equation}
where $\mathbf{U} = \sum_{l} U_{l}$ is the displacement operator labeled by the degree of the cochain it acts on.
 
It is immediate that $\mathbf{U}$ exponentiates a zero form with value in the representation of the Lie algebra. It seems at first that we are 
coupling $(n-p)$-form fields to $p$-fields. However it is not the case, as it should be physically. The cochains cannot be used as the basic fields in the theory, since the gauge invariance of the resulting extended objects is not well-defined, as illustrated by the no-go theorem of Teitelboim \cite{teitelboim}.
 Thus, although we use inhomogeneous differential forms to represent the algebra of spinors, they exist at points within a cell. 
This argument indicates that one may not attempt to make the $SU(4)$ symmetry local in our method as opposed to the suggestion of \cite{BJ}. Nevertheless, our action has the discrete rotational symmetry within a given cubic cell.

Let us look at the 2D example and consider a square as below. 
\begin{figure}[H]
\label{fig:V}
{\par\centering \includegraphics{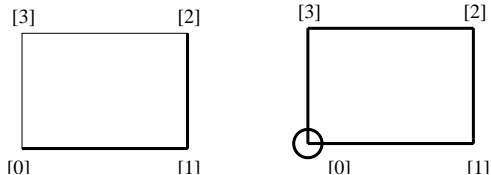} \par}
\caption{Two types of couplings: two edges with a vertex in common and the entire square with one vertex.}
\end{figure} 
The displacement $\mathbf{U}$ in the group then corresponds to matching fields at a fixed point in each cell.

To summarise \footnote{One of the present authors; V de B, intends to address in more detail the issue of gauging in subsequent work.}, in gauging the fermion field, we insert an operator to the right of the wedge in the action. In our language, the operator $d- \delta$ affects a displacement and so one should propagate the field by inserting the appropriate $\mathbf{U}$ operator.

Finally, a short comment about fermion masses. It has been established by various authors \cite{Kovacs, mitra} that the introduction of masses in the theory, either by hand as specifying bare masses or those induced by renormalization counter terms will depend on the gauging procedure. As pointed out above, we do not block the fields in the present method and it has been shown \cite{thun} that for the usual Susskind fermions with parallel transport operators inserted in the Lagrangian, no counterterms arise from renormalization and so the bare masses specified by hand are the actual masses in the theory. This means that our mass term which is

\begin{equation}
m \bar{\Phi} \wedge \star \Phi,
\end{equation}
would give rise to no counterterms.
In the usual methods, the blocking of the fields is central as the residual doubling present in these methods \cite{BJ, golterman} is identified as providing the multiple flavours. Moreover, careful analysis of the various symmetries of the Lagrangian \cite{Kovacs, mitra} indicate that the DK lattice theory and the staggered theory differ as regards the generation of masses. For instance, the shift invariance is present in the staggered formalism thus preventing the dynamical generation of masses while it is not in DK after gauging. 

\section{A crucial remark}

We have made some general comments about gauging, which ought to be the crucial step; but it is essential to note that this theory can be related to the staggered theory in the standard way of Becher and Joos, where one would expect the problem of doubling to reappear in the form of species doubling. What is novel in our approach is that by keeping inhomogeneous cochain fields without mapping their coefficients to vertices as is customary, we open the possibility of having reduced Dirac-K\"ahler fermions regularized in a way that has fields not defined at vertices solely but also on links and plaquettes. This is a consequence of the close relation of our formulation to differential geometry which makes the shift operators of FIG. 1 superfluous. Yet the extended objects should have no intrinsic meaning physically and only represent a fermionic field $\psi$. 

This state of affairs is clearly different from the type of theories covered by the Nielsen-Ninomiya theorem \footnote{V de B would like to thank Noboru Kawamoto for some discussions of these issues at Lattice 2003.}, yet the theory is discretised so we expect to be able to put it on a computer.

\section{Conclusion}
We are thus able to work with chiral Dirac fermions without 
seemingly suffering from degeneracy. In summary:
\begin{itemize}
\item Spectral degeneracy does not occur in the Dirac-K\"ahler formalism,
\item Species doubling is avoided using the GD wedge which allows us to
impose the subsidiary conditions,
\item Species doubling is also avoided since we have $\delta=\star d\star $ and $\star\star=I$;
the DKE is overdetermined if this is not the case,
\item Chirality is captured since we have a discrete Hodge star.
\item Some gauging prescriptions are made for the resulting staggered fermion theory and we have alluded to a new approach which is more natural in the present context.
\end{itemize}

We have tried to keep this note as short as possible, avoiding purposely to 
cloud the argument with the notation used in GD. Yet we intend to write a more 
exhaustive note which will include more technical details of our method (GD).
The interest for this problem was initiated partly by the paper of Rabin in 
which the role of the Hodge star is presented as crucial to the lattice 
formulation of Chiral fermions. We also hope to have conveyed the idea that 
a large class of discretisation problems are caused by discrepancies in the 
relations between the operators $\wedge, \; d$ and $\star$. This is precisely 
what we tried to remedy here, emphasising the importance
of differential geometric ideas in a discretised theory.

\begin{acknowledgments}
We would like to thank James Sexton for drawing our attention to applications of
geometric discretisation to fermions and introducing  
references \cite{BJ, rabin} to us. Thanks also to Denjoe O'Connor and Siddhartha Sen for stimulating discussions. We are grateful to our referees for bringing \cite{Kovacs, mitra} to our attention.  V de B was supported by the HEA through the IITAC. 
\end{acknowledgments}

\end{document}